\documentclass[journal=jacsat,manuscript=article,layout=twocolumn]{achemso}

\usepackage{chemformula} % Formula subscripts using \ch{}
\usepackage[T1]{fontenc} % Use modern font encodings
\usepackage{amssymb}
\usepackage{caption}
\author{Mark Kamper Svendsen}
\affiliation[DTU]
{Department of Physics, Technical University of Denmark, Fysikvej, DK-2800 Kongens Lyngby, Denmark}

\author{Hiroshi Sugimoto}
\affiliation[Kobe]
{Department of Electrical and Electronic Engineering, Kobe University, Rokkodai, Nada, Kobe 657-8501, Japan}

\author{Artyom Assadillayev}
\affiliation[DTU]
{Department of Physics, Technical University of Denmark, Fysikvej, DK-2800 Kongens Lyngby, Denmark}

\author{Daisuke Shima}
\affiliation[Kobe]
{Department of Electrical and Electronic Engineering, Kobe University, Rokkodai, Nada, Kobe 657-8501, Japan}

\author{Minoru Fujii}
\affiliation[Kobe]
{Department of Electrical and Electronic Engineering, Kobe University, Rokkodai, Nada, Kobe 657-8501, Japan}

\author{Kristian S. Thygesen}
\affiliation[DTU]
{Department of Physics, Technical University of Denmark, Fysikvej, DK-2800 Kongens Lyngby, Denmark}
\email{thygesen@fysik.dtu.dk}

\author{S\o ren Raza}
\affiliation[DTU]
{Department of Physics, Technical University of Denmark, Fysikvej, DK-2800 Kongens Lyngby, Denmark}
\email{sraz@dtu.dk}

\title{Ultraviolet Mie resonances in computationally discovered boron phosphide 
nanoparticles}

\begin{document}

%\begin{tocentry}

%\end{tocentry}

\begin{abstract}
Controlling ultraviolet light at the nanoscale using optical Mie resonances holds great promise for a diverse set of applications, such as lithography, sterilization, and biospectroscopy. However, Mie resonances hosted by dielectric nanoantennas are difficult to realize at ultraviolet wavelengths due to the lack of both suitable materials and fabrication methods. Here, we systematically search for improved materials by computing the frequency dependent optical permittivity of 338 binary semiconductors and insulators from first principles, and evaluate their potential performance as high refractive index materials using Mie theory. Our analysis reveals several interesting candidate materials among which boron phosphide (BP) appears particularly promising. We then prepare BP nanoparticles and demonstrate that they support Mie resonances at visible and ultraviolet wavelengths using both far-field optical measurements and near-field electron energy-loss spectroscopy. We also present a laser reshaping method to realize spherical Mie-resonant BP nanoparticles. With a refractive index above 3 and low absorption losses, BP nanostructures advance Mie optics to the ultraviolet.
\end{abstract}

\section*{Introduction}
Achieving control over ultraviolet light with nanoscale materials is essential for improving surface-enhanced spectroscopies of biological molecules and enabling new ultraviolet optical components. Geometric Mie resonances supported by resonant nanoantennas made from materials that combine a high refractive index with low absorption losses offer efficient and tunable manipulation of the near- and far-field of optical waves\cite{Kuznetsov2016,Kruk2017}. Mie resonances have been realized at visible and infrared wavelengths thanks to the mature lithographic processing of suitable materials\cite{baranov2017all}, such as silicon\cite{Staude2017}, gallium phosphide\cite{Wilson2020}, and titanium dioxide\cite{Chen2018}. It would be desirable to extend the operation of these materials to the ultraviolet, but their small direct band gap energies ($\lesssim3$~eV) lead to significant absorption losses in the ultraviolet. Recently, metasurfaces composed of an array of nanostructured materials with a wide band gap and moderate refractive index ($n\approx2.1-2.3$) have demonstrated wave front manipulation in the ultraviolet using waveguide modes\cite{Huang2019,Zhang2020,Zhao2021}. For ultraviolet Mie optics, diamond has been theoretically suggested as a potential material~\cite{Gutierrez2018,Hu2020}, but diamond comes with significant nanofabrication challenges\cite{Aharonovich2011}. Consequently, extending the rich optical properties of Mie resonances observed in the visible to the ultraviolet requires identification of new high-index materials as well as development of suitable fabrication methods to realize Mie-resonant nanoantennas. 

Concurrent advances in first-principles methodology and computing power have recently made it possible to design and discover new materials via high-throughput computations\cite{curtarolo2013high,jain2013commentary,kirklin2015open,c2db1,c2db2}. The approach has been successfully applied in several domains, including photovoltaics, transparent conductors, and photocatalysis\cite{BP_transparant_conductor_HT,yu2012identification,castelli2012computational}. However, to the best of our knowledge, computational discovery of new high-index materials remains largely unexplored. Relevant previous work in this direction has been limited to the static response regime\cite{petousis2017high,HT_refractive_index} reflecting the fact that the major materials databases so far has focused on ground state properties. 

Here we use high-throughput linear response density functional theory (DFT) to screen an initial set of 2743 elementary and binary materials with the aim to identify isotropic high-index, low loss, and broad band optical materials. For the most promising materials, the computed frequency-dependent complex refractive indices are used as input for Mie scattering calculations to evaluate their optical performance. In addition to the already known high-index materials we identify several new compounds. In particular, boron phosphide (BP) offers a refractive index above 3 with very low absorption losses in a spectral range spanning from the infrared to the ultraviolet. We then prepare BP nanoparticles and show, by means of dark-field optical measurements and electron energy-loss spectroscopy, that they support size-dependent Mie resonances in the visible and ultraviolet. Finally, we demonstrate a laser reshaping method to realize spherical BP nanoparticles, which host multiple Mie resonances in quantitative agreement with full-field optical simulations. 

\section*{Results}

\subsection*{High-throughput screening}

\begin{figure}
    \centering
    \includegraphics[width=0.45\textwidth]{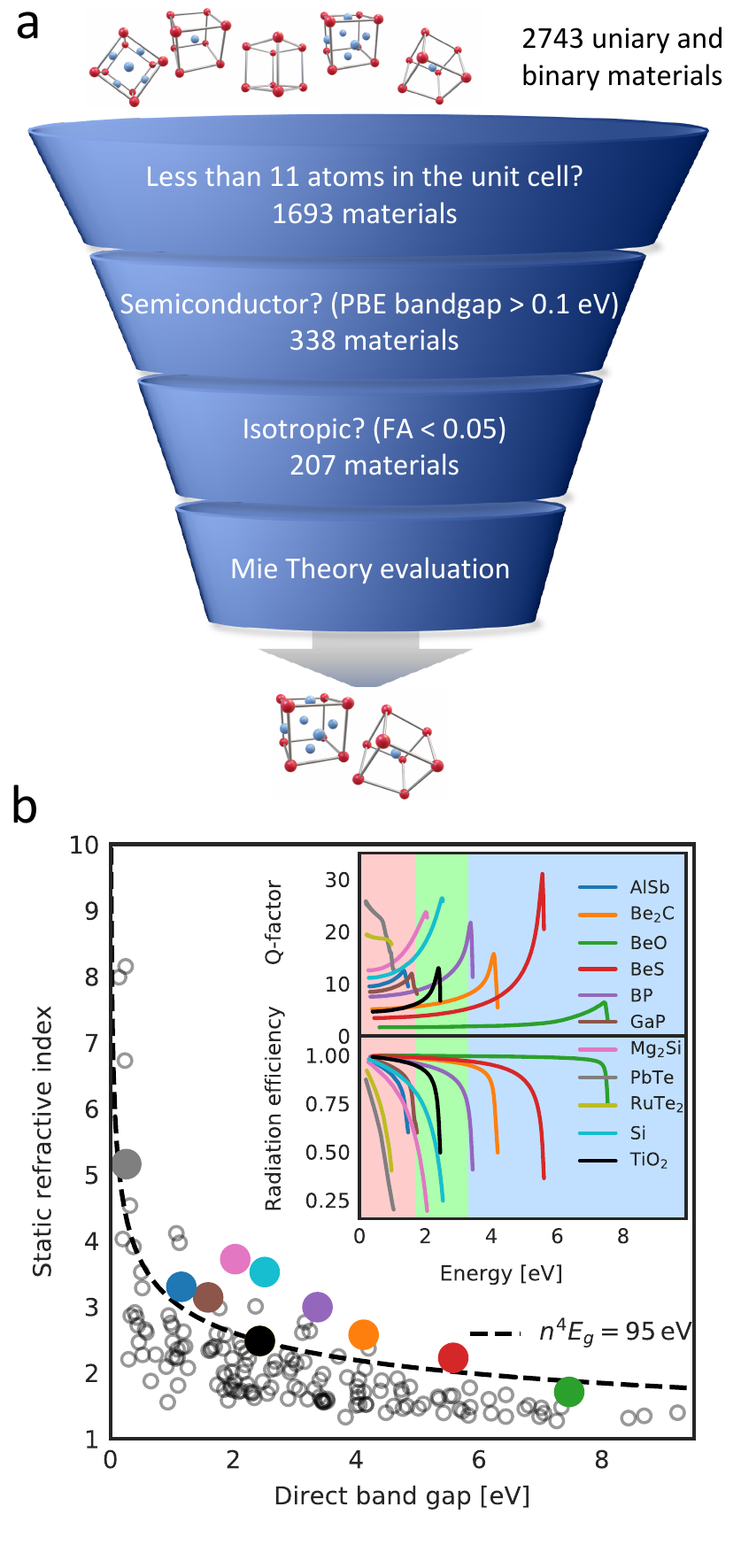}
    \caption{\textbf{High-throughput materials screening.} \textbf{a,} Schematic of the screening steps and the number of materials that survive each of them. \textbf{b,} The static refractive index of the isotropic materials plotted as a function of their direct band gap energies along with the Moss formula. The colored dots highlight interesting materials and the inset shows the energy-dependent $Q$-factor and radiation efficiency, $\eta$, of the MD resonances in those materials. The background colors represent the infrared (red), visible (green) and ultraviolet (blue) spectral regions.}
    \label{fig:fig1}
\end{figure}

Our high-throughput screening procedure is illustrated in Fig.~\ref{fig:fig1}a. We build the workflow using the Python-based Atomic Simulation Recipes (ASR)\cite{gjerding2021atomic} framework and the MyQueue scheduling software\cite{mortensen2020myqueue} (see detailed workflow in Supplementary~Note~1 and Supplementary~Fig.~1). Starting from 2743 thermodynamically stable elementary and binary materials from the Open Quantum Materials Database (OQMD)\cite{kirklin2015open} we extract the 1693 materials with up to 10 atoms in the unit cell and relax the atomic structure using DFT with the Perdew-Burke-Ernzerhof (PBE) exchange correlation functional\cite{PBE} and the D3 correction to account for the van der Waals forces\cite{DFTD3}. We perform DFT ground state calculations for all of the materials to determine their electronic band gaps. After discarding the metals, we are left with 338 semiconductors for which we calculate the optical dielectric function, $\epsilon(\omega)$, within the random phase approximation (RPA) and extract the refractive index and extinction coefficient as $n(\omega) = \mathrm{Re}\left\{\sqrt{\varepsilon(\omega)}\right\}$ and $k(\omega) = \mathrm{Im}\left\{\sqrt{\varepsilon(\omega)}\right\}$, respectively. All calculations are performed with the GPAW code\cite{enkovaara2010electronic,yan2011linear}. 

Next, we classify the materials according to the anisotropy of their refractive index tensor using a cut-off of 0.05 for the fractional anisotropy (see Supplementary~Fig.~2). This leaves us with 207 isotropic material candidates, for which we show the static refractive index as a function of the direct band gap in Fig.~\ref{fig:fig1}b. The data points qualitatively follow the Moss formula\cite{moss1985relations} (dashed line in Fig.~\ref{fig:fig1}b); however, there are significant deviations from the general trend, which we ascribe to variations in oscillator strength and density of the transitions across the direct gap. It is well known that (semi)local functionals like the PBE employed in the present work systematically underestimates band gaps\cite{perdew1985density}. This effect is, however, to some extent compensated by the fact that the RPA neglects the attractive electron-hole interactions and consequently underestimates the spectral weight near the band edge. As a result, refractive indices obtained with RPA@PBE are typically in good agreement with experiments in the static limit\cite{yan2011linear} while deviations occur at higher frequencies near the band edge region (see Supplementary~Note~2 and Supplementary~Fig.~3-5). We shall return to this point later, but for now mention that for the materials that are identified as interesting based on this screening we will employ more accurate and computational expensive many-body perturbation methods.

We now turn to a more in-depth evaluation of the performance of the discovered materials. Specifically, we use Mie theory to calculate the scattering properties of a spherical nanoparticle made from the subset of isotropic materials. We focus on the lowest-order magnetic dipole (MD) resonance of the spheres and calculate energy-dependent quality factors, $Q$, and radiation efficiencies, $\eta$ (see Supplementary~Note~1)\cite{baranov2017all}. This is achieved by continuously adjusting the size of the sphere to tune the energy of the MD resonance across the infrared, visible and ultraviolet regions. A high $Q$-factor is beneficial for boosting the local field enhancement of the nanoparticle, while a radiation efficiency close to unity points to low absorption losses. This analysis identifies all of the commonly used materials, such as Si, TiO$_2$, and GaP. However, we also find a number of other highly promising materials, some of which are highlighted in the inset of Fig.~\ref{fig:fig1}b. In particular, we identify BP, which has a refractive index exceeding that of TiO$_2$ and a radiation efficiency higher than both silicon and TiO$_2$ across the entire visible part of the spectrum. For these reasons, we believe BP stands out as an overlooked material with highly desirable optical properties and we will focus on BP in the rest of the paper. 

\subsection*{Optical response of BP}\label{sec:optical}

\begin{figure}
    \centering
    \includegraphics{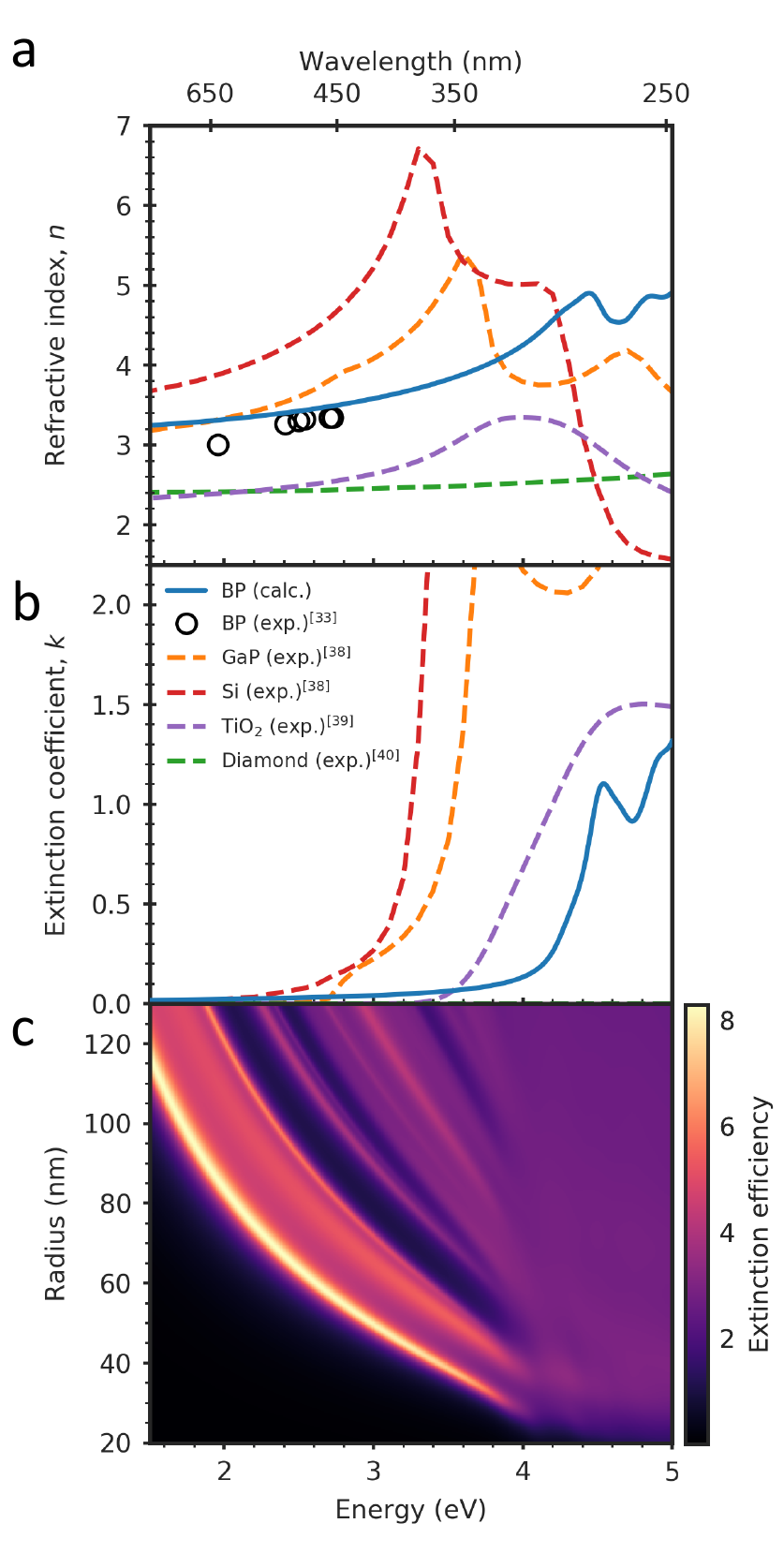}
    \caption{\textbf{Optical response and Mie resonances of BP.} \textbf{a,b} Refractive index $n$ and extinction coefficient $k$ of BP calculated using the f-sum corrected BSE method compared with that of some of the commonly used dielectrics in the visible and ultraviolet spectral regions. \textbf{c,} Extinction efficiency map of BP spheres of varying radii calculated using Mie theory, which demonstrates that Mie resonances can be sustained in the visible and ultraviolet.}
    \label{fig:fig2}
\end{figure}

BP crystals were successfully synthesized as early as 1957\cite{Popper1957}, yet experimental measurements of its refractive index are limited to a couple of data points in the visible\cite{Takenaka1976,Wettling1984}. Refractive index measurements of BP thin films have also been conducted but with varying results\cite{Odawara2005,Dalui2008}. 

The RPA@PBE permittivities used for the initial screening are qualitatively accurate but suffer from underestimated band gaps and missing excitonic effects. To determine the refractive index of BP with quantitative accuracy, we solve the Bethe-Salpeter equation (BSE) to obtain the permittivity using single-particle transition energies obtained from a G$_0$W$_0$ band structure calculation. The band structure calculation reveals an indirect band gap of 2.1~eV and a direct band gap of 4.41~eV, which matches experimental measurements of the band gap energies for BP.\cite{Archer1964,Schroten1998} The square root dependence of the refractive index on the permittivity makes it crucial to converge both the real and imaginary parts of the latter. Unfortunately, the real part converges slowly with the number of bands making it impractical to obtain well converged results directly from the BSE. The problem can be alleviated by extending the imaginary part of the permittivity by an exponentially decaying tail whose weight is fixed by the f-sum rule (see Methods), and subsequently obtain the real part via the Kramers--Kronig relation. We benchmark this approach against experimental data for the refractive index of crystalline silicon and find excellent agreement (see Supplementary~Note~3 and Supplementary~Fig.~6). 

With the f-sum rule fulfilling BSE-G$_0$W$_0$ method at hand, we are in a position to make a quantitative comparison of the refractive index of BP with some of the commonly used materials\cite{aspnes1983dielectric,siefke2016materials} in the visible as well as diamond\cite{phillip1964kramers}, which has been theoretically suggested for operation in the ultraviolet (Fig.~\ref{fig:fig2}a,b). We observe that the absorption edge of BP lies significantly higher than silicon, GaP, and TiO$_2$, while it retains a refractive index comparable to that of GaP. This suggests that BP provides low-loss operation across the entire visible spectrum. While this is also the case for TiO$_2$, its refractive index is significantly lower than that of BP. The higher refractive index of BP means that nanostructures can be made more compact\cite{Yang2017} and packed more densely for enhanced metasurface performance\cite{Lalanne2017}. Importantly, we note that BP offers a high refractive index with a low extinction coefficient not only in the visible but also in the ultraviolet -- a spectral region which is unreachable with the commonly-used materials. Diamond is transparent in the ultraviolet as well, but has a significantly lower refractive index. We illustrate the broadband performance of BP by performing extinction efficiency calculations using Mie theory for a BP sphere with varying radii, confirming that Mie resonances can be sustained across the visible and ultraviolet (Fig.~\ref{fig:fig2}c).

\begin{figure*}[!t]
    \centering
    \includegraphics[width=\textwidth]{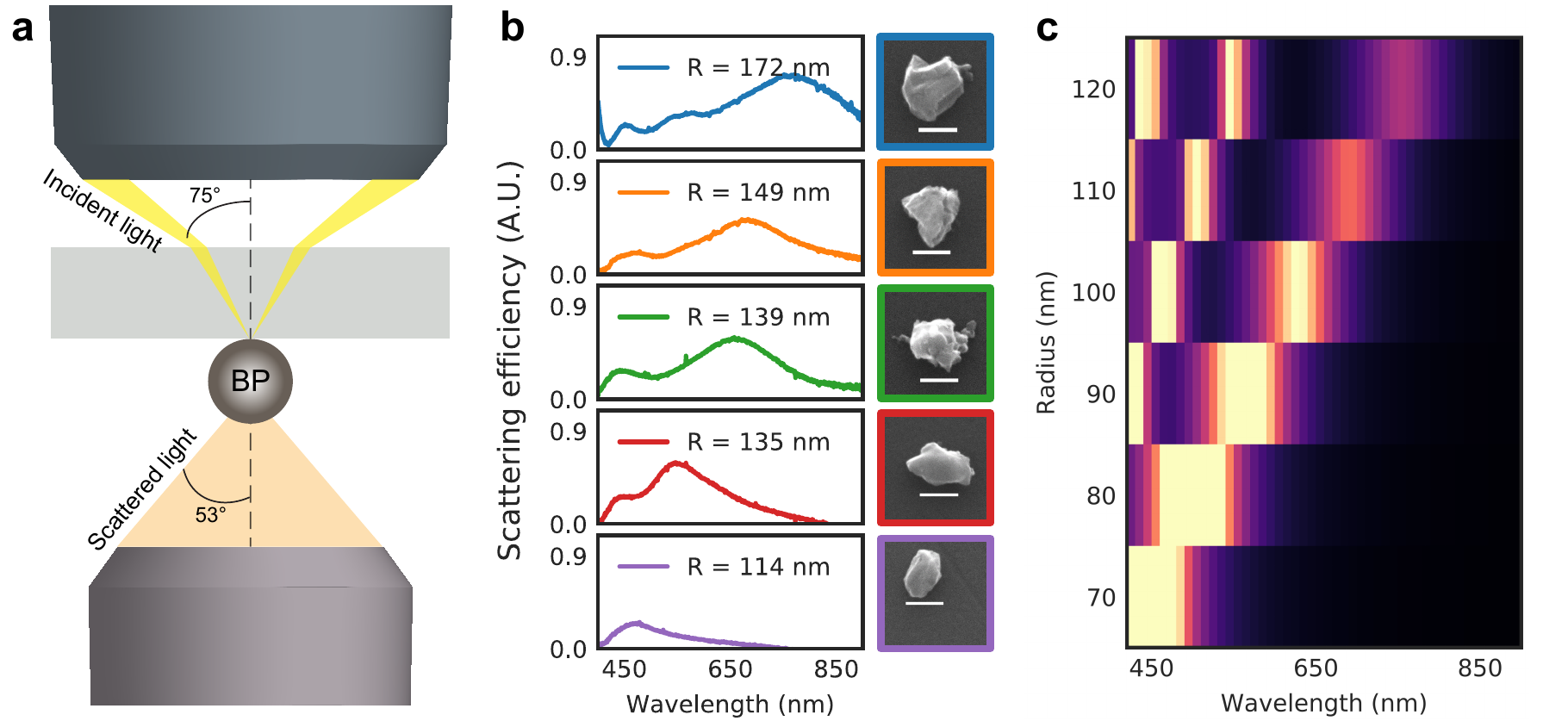}
    \caption{\textbf{Dark-field scattering of BP nanoparticles.} \textbf{a,} Schematic of the dark-field scattering spectroscopy. The nanoparticles are excited by unpolarized, white light incident at an oblique angle and the transmitted scattered light is collected. \textbf{b,} Scattering efficiency spectra recorded from individual BP nanoparticles of different sizes showing clear scattering peaks due to resonant interaction. The effective radius $R$ of the nanoparticle is determined from the area of the particles, assuming a spherical shape. \textbf{c,} Simulated scattering efficiency map of BP nanospheres with varying particle radii.}
    \label{fig:fig3}
\end{figure*}

\subsection*{Far- and near-field characterization of BP nanoparticles}

We now turn to an experimental demonstration of the potential of BP as a Mie-resonant nanostructure. BP has been synthesized in various forms such as crystals\cite{Kumashiro1990,Woo2016}, films\cite{Dalui2008,Kumashiro2011,Padavala2016}, and nanoparticles\cite{Sugimoto2015,Feng2019,Sugimoto2019}. However, BP nanoparticles in the size range suitable for sustaining Mie resonances have not been reported. We prepare BP nanoparticles in the size range of a few hundred nanometers by grounding BP powder (Kojundo Chemicals) in a mortar and then dispersing it in methanol. The BP solution is subsequently dropcasted on a glass substrate.

We measure the far-field scattering efficiency of individual nanoparticles by illuminating the nanoparticles through a high-numerical-aperture dark-field objective and collecting the transmitted scattered light (see Fig.~\ref{fig:fig3}a). A similar measurement setup has been used to detect Mie resonances in silicon nanoparticles.\cite{Sugimoto2017} The scattering spectra recorded from a series of BP nanoparticles clearly show resonant peaks, which red shift with increasing particle size (Fig.~\ref{fig:fig3}b). Despite the irregular particle shapes, the scattering resonances are quite prominent and follow the trend observed in other Mie-resonant nanostructures, namely, that the smallest particle size support only the lowest-order Mie resonance while larger particle sizes also support higher-order Mie resonances\cite{Sugimoto2020}. We additionally confirm the crystallinity of the nanoparticles using micro-Raman spectroscopy from individual nanoparticles (see Supplementary~Fig.~7). The particle radii are extracted from the scanning electron microscopy images under the assumption of a spherical shape and used as input for full-field simulations of the scattering efficiency of a BP nanosphere. The simulations account for the measurement setup as well as the glass substrate (see Methods). We find that the simulations accurately reproduce both the shift in resonance wavelengths with particle size as well as the number of resonant peaks (Fig.~\ref{fig:fig3}c). However, the particle sizes need to be adjusted to match the resonance wavelengths observed in the experiments. This suggests that the particle shape is better characterized as flakes with a thickness significantly smaller than the in-plane size. Nonetheless, the distinctive, multiple scattering peaks provide strong evidence for the interpretation that these are related to geometric Mie resonances.

\begin{figure}[!t]
    \centering
    \includegraphics{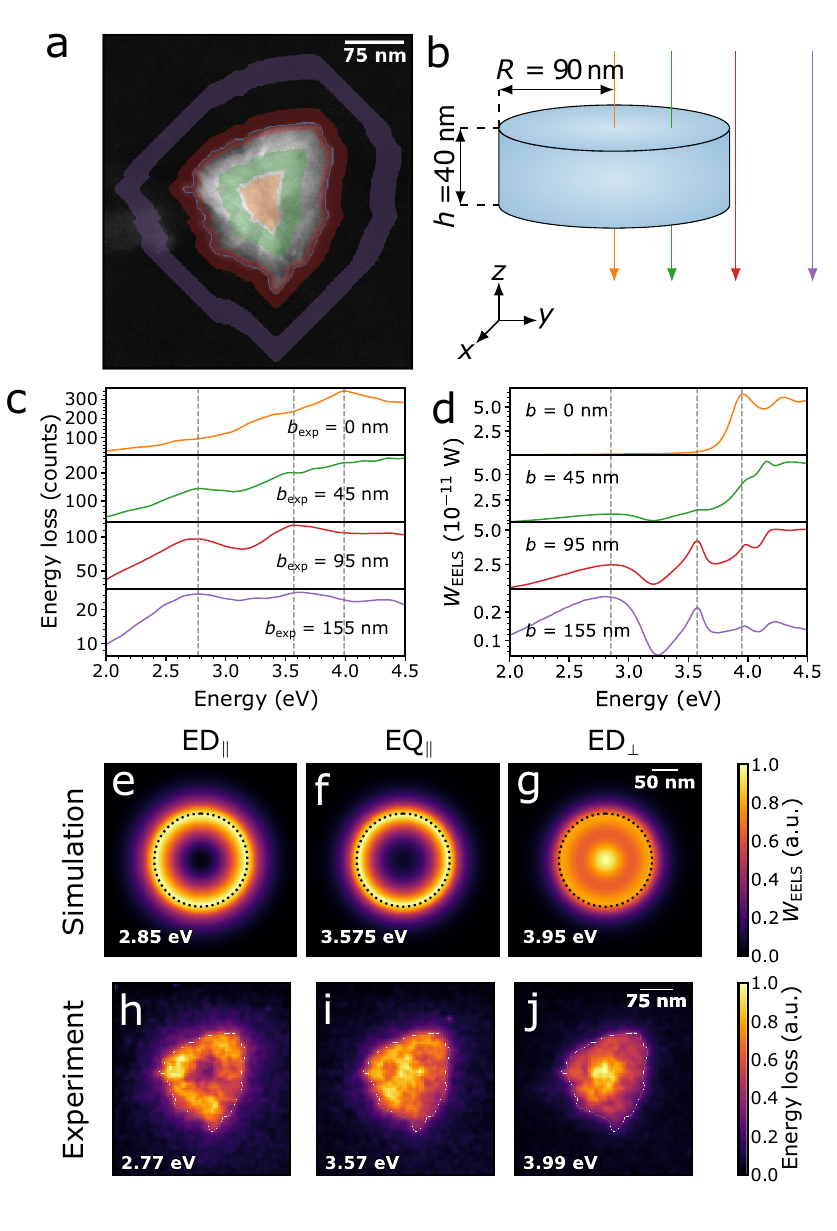}
    \caption{\textbf{Electron energy-loss spectroscopy of BP nanoparticles.} \textbf{a,b} Experimental STEM image and theoretical set-up depicting the electron beam positions $b$ used to acquire the measured and simulated EELS spectra of \textbf{c,d}, respectively. The triangular-shaped BP nanoparticle is characterized by an effective radius of $R=90$~nm. Several EELS peaks due to Mie resonances are observed. \textbf{e-g} Simulated and \textbf{h-j} experimental EELS intensity maps at the Mie resonance energies. The Mie modes are identified as the in-plane electric dipole (ED$_\parallel$), in-plane electric quadrupole (EQ$_\parallel$), and out-of-plane electric dipole (ED$_\perp$) using a multipole decomposition.}
    \label{fig:fig4}
\end{figure}

To gain more insight into the nature of these resonances and to access the ultraviolet spectral region, we also perform near-field characterization on similar BP nanoparticles using electron energy-loss spectroscopy (EELS). EELS is performed in a transmission electron microscope and has been employed to access near-field properties of both metallic\cite{GarciaDeAbajo2010,Colliex2016} and dielectric nanostructures\cite{Assadillayev2021,Assadillayev2021b,Alexander2021} as well as optical devices\cite{Song2021}. The combined high spatial and spectral resolution of EELS provides unique nanoscale information on optical modes over a broad spectral range. For EELS measurements, the BP nanoparticles are deposited on a thin silicon oxide membrane. The EELS signal recorded from a triangular-shaped BP nanoparticle (Fig.~\ref{fig:fig4}a) at different beam positions is presented in Fig.~\ref{fig:fig4}c. The position of the beam is directly related to the excitation efficiency of the optical modes~\cite{Raza2016}, and thus, by judicious positioning of the beam we can selectively excite different Mie modes.~\cite{Assadillayev2021} When the beam is positioned in the center of nanoparticle, we observe a distinct resonance in the ultraviolet at 3.99~eV. As the beam is moved closer to the surface of the nanoparticle, two additional resonances are observed at the energies 3.57~eV and 2.77~eV. To identify the nature of these resonances, we simulate the EELS signal of a BP nanodisk with the same effective radius as the measured BP nanoparticle (Fig.~\ref{fig:fig4}b). The thickness of the nanodisk is varied to achieve correspondence to the measured EELS resonance energies. The simulated EELS spectra for a nanodisk thickness of $h=40$~nm at the same beam positions as in the experiments are shown in Fig.~\ref{fig:fig4}d. Here, we observe that the three lowest-energy EELS peaks have the same dependence on the impact parameter of the electron beam as seen experimentally. The simulated resonance energies of all three EELS peaks are also in quantitative agreement with the experiments, albeit the lowest-energy EELS peak is slightly shifted to higher energies in the simulations. The measured EELS peaks are broadened by the finite energy resolution of our EELS setup (see Methods). By performing a multipole decomposition of the induced field produced by the electron beam\cite{Alaee2018}, we identify the two lowest energy EELS peaks to be due to the Mie modes of the in-plane electric dipole ED$_\parallel$ and in-plane electric quadrupole EQ$_\parallel$. The highest energy EELS peak has contributions from both the in-plane magnetic dipole and and out-of-plane electric dipole ED$_\perp$, where the latter dominates in the center of the particle (see Supplementary~Fig.~8 for full decomposition). Simulated EELS intensity maps show that the electron beam couples efficiently to the in-plane modes, ED$_\parallel$ and EQ$_\parallel$ for beam positions near the surface of the particle, while the out-of-plane ED$_\perp$ is excited also for beam positions in the center of the particle (Fig.~\ref{fig:fig4}e-g). The experimental EELS intensity maps of these three Mie modes are in good agreement with the simulations (Fig.~\ref{fig:fig4}h-j) as well as previous EELS measurements performed on Mie-resonant silicon nanoparticles\cite{Assadillayev2021,Assadillayev2021b}. The near-field EELS measurements along with far-field dark-field optical measurements demonstrate that, despite their irregular shape, BP nanoparticles host a variety of multipolar size-dependent Mie resonances across the visible and ultraviolet, which are key attributes of low-loss high-index nanostructures.

\subsection*{Laser reshaping}
\begin{figure}
    \centering
    \includegraphics[width = 0.9\columnwidth]{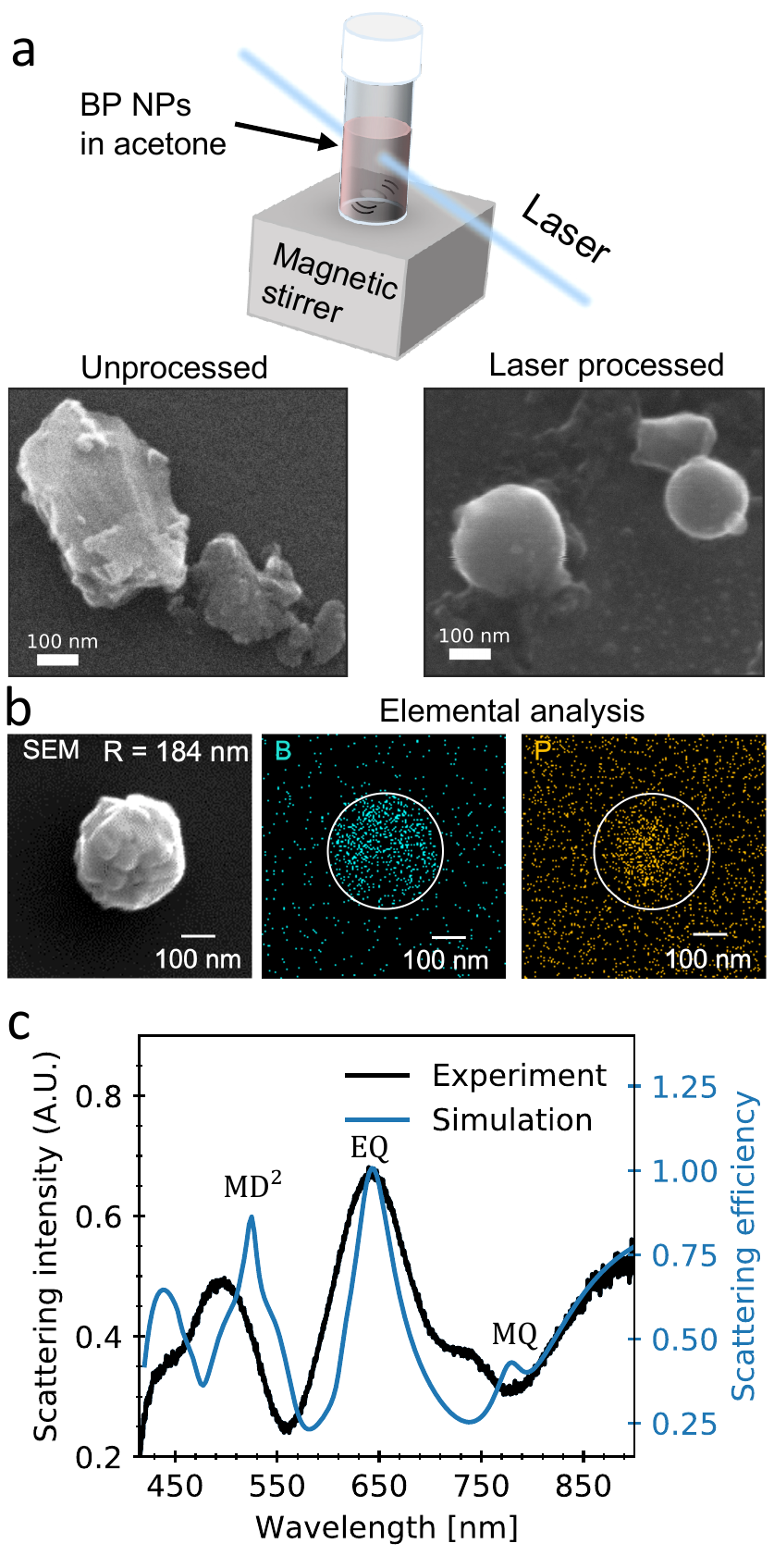}
    \caption{\textbf{Laser reshaping of BP nanoparticles.} \textbf{a,} Reshaping of the BP nanoparticles is achieved by pulsed laser irradiation of unprocessed BP nanoparticles in acetone. After the irradiation, the BP nanoparticles obtain a spherical shape. \textbf{b,} Scanning electron microscopy image and energy-dispersive x-ray analysis of reshaped BP nanoparticle, showing a spherical shape without change in composition. \textbf{c,} Scattering spectrum recorded from the BP nanoparticle in \textbf{b} and simulated spectrum for a BP nanosphere with radius $R=165$~nm with multipole decomposition revealing the excitation of the magnetic quadrupole (MQ), electric quadrupole (EQ), and radial higher-order magnetic dipole (MD$^2$).}
    \label{fig:fig5}
\end{figure}

To alleviate the irregular shape of the as-prepared BP nanoparticles, we generate spherical BP nanoparticles by irradiating the unprocessed BP nanoparticles with a pulsed laser (see Fig.~\ref{fig:fig5}a and Methods). An example of a laser-processed BP nanoparticle is presented in Fig.~\ref{fig:fig5}b, confirming that the laser processing can be used for realizing spherical-shaped BP nanoparticles without affecting the nanoparticle composition. In addition, we perform micro-Raman spectroscopy on the same particle and find that the crystallinity from the unprocessed BP nanoparticles is retained after laser reshaping (see Supplementary~Fig.~7). The dark-field scattering spectrum from the BP nanoparticle in Fig.~\ref{fig:fig5}b is recorded and we observe multiple scattering peaks (Fig.~\ref{fig:fig5}c). The relatively large particle radius ($R=184$~nm) places the lowest-order Mie resonances at wavelengths longer than our measurement range, while the scattering peaks observed can be attributed to higher-order Mie resonances. The full-field simulation of a BP nanosphere with a slightly smaller radius $R=165$~nm, where we account for the substrate and the measurement setup, shows very good agreement with the measurement. We attribute the deviation in particle radius to shape imperfections and a slight variation between the calculated and experimental refractive index of BP. Multipole decomposition reveals that the scattering peaks are due to the excitation of the magnetic quadrupole MQ, the electric quadrupole EQ, and radial higher-order magnetic dipole MD$^2$, thereby confirming the Mie-resonant nature of BP nanoparticles. We also performed EELS measurements on a smaller laser-reshaped BP nanoparticle, where we observe Mie resonances in the ultraviolet (see Supplementary~Fig.~9).

\section*{Discussion}
Using a DFT-based high-throughput screening method combined with optical Mie theory of 338 dielectrics, we identify BP as a promising high-refractive-index material for ultraviolet nanooptics. We develop a new, quantitatively accurate many-body perturbation theory based methodology for calculating refractive indices and use it to reveal that BP has a high refractive index ($n>3$) and low extinction coefficient ($k<0.1$) up to ultraviolet photon energies of 4~eV. We present an approach to fabricate BP nanoparticles as well as a laser reshaping method to generate spherical BP nanoparticles. Through dark-field optical spectroscopy and EELS measurements, we confirm the presence of Mie resonances in BP nanoparticles across the visible and ultraviolet. Our work advances nanoscale Mie optics to the ultraviolet and may find applications in metasurface-enhanced spectroscopy of biological molecules and, more generally, in realizing metasurface optical components operating in the ultraviolet.

\section*{Methods}

\subsection*{Computational workflow}
\paragraph*{Structural relaxation:}
All ground- and excited state calculations were performed with the GPAW electronic structure  code\cite{enkovaara2010electronic}.
We relaxed the atomic structure and the unit cell of the materials until the maximum force (stress) is below $10^{-4}$ eV/Å (0.002eV/Å$^3$). We use the PBE functional for exchange and correlation effects, a $\Gamma$-point centered k-point grid with a density of $6.0$Å$^{-3}$, a 800 eV plane wave cutoff and a Fermi-Dirac smearing of 50 meV. Van der Waals interactions were taken into account by the D3 correction scheme\cite{DFTD3}. 

\paragraph*{RPA calculations:}
We calculate the optical permittivity, $\varepsilon(\omega)$, within the Random Phase Approximation(RPA) using the dielectric function module in GPAW. From $\epsilon(\omega)$ we calculate the refractive index and extinction coefficient as $n(\omega) = \mathrm{Re}\left\{\sqrt{\varepsilon(\omega)}\right\}$ and $k(\omega) = \mathrm{Im}\left\{\sqrt{\varepsilon(\omega)}\right\}$ respectively. To ensure convergence across all materials we employ a k-point grid with a high density of 20.0 Å$^{-3}$ and include conduction bands up to 5 times the number of valence bands. The calculations were performed on a nonlinear frequency grid with an initial frequency spacing of 0.5 meV, a broadening of 50 meV and a local field cutoff of 50 eV. \\

\subsection*{Many-body perturbation theory calculations}
\paragraph*{G$_0$W$_0$ calculations:} The G$_0$W$_0$ calculations were performed on top of the ground state calculations. To ensure converged quasi-particle gaps we perform extrapolation of both the plane wave cut-off and the k-point resolution to infinity. 

\paragraph*{BSE calculations:} The BSE calculations were performed within the Tamm-Dancoff approximation on a k-point grid with a density of 20 Å$^{-3}$. The calculation included all valence and conduction bands within 2.3 eV of the valence band maximum and conduction minimum, respectively. The calculation of the screened interaction included all occupied bands and unoccupied bands up to 5 times the number of occupied bands, and we accounted for local field effects up to a plane wave cut-off of 50 eV. The calculation was performed on a linear 10001 point frequency grid spanning 0 to 8 eV.

\paragraph*{BSE f-sum rule correction}
The optical polarizability has to obey the f-sum rule,
\begin{align}\label{eq:chi_sum_rule}
    \int_0^\infty d\omega \omega \mathrm{Im}\chi(\omega) = \frac{\pi}{2}\frac{n_ee^2}{m}\,, 
\end{align}
\noindent where $n_e$ is the electron density, $e$ is the elementary charge and $m$ is the electron mass. Since it is not obvious what to use for $n_e$ in our PAW calculations, we use a different strategy to fix the left hand side, namely we obtain it from an RPA calculation (which can be readily converged) 
\begin{align}\label{eq:chi_sum_rule_area_RPA_andBSE}
    \int_0^\infty d\omega \omega \mathrm{Im}\chi^{\mathrm{(RPA)}}(\omega) = \int_0^\infty d\omega \omega \mathrm{Im}\chi^{\mathrm{(BSE)}}(\omega)\,.
\end{align}
\noindent Eq. \ref{eq:chi_sum_rule_area_RPA_andBSE} can be enforced if we artificially extend the imaginary part of the BSE polarizability. Denoting the largest transition energy included in the BSE calculation as $w_c$, we perform the following extension,
\begin{align}\label{eq:extention_imBSE}
    \mathrm{Im}\chi^{\mathrm{(BSE)}}(\omega) = 
    \begin{cases}
            \mathrm{Im}\chi^{\mathrm{(BSE)}}(\omega), &         \text{if } \omega \leq \omega_c,\\
            C_0e^{-\gamma(\omega-\omega_c)}, &         \text{if } \omega > \omega_c.
    \end{cases}
\end{align}
The constant $C_0$ is used to ensure continuity and the constant $\gamma$ is fixed to give the correct spectral weight as fixed by Eq. \ref{eq:chi_sum_rule_area_RPA_andBSE}. For benchmarks see Supplementary Information.

\subsection*{Optical simulations}
The scattering efficiency and EELS simulations are both performed in COMSOL Multiphysics (ver. 5.6), which solves Maxwell's equations using finite-element modelling. For the scattering efficiency simulations in Fig.~\ref{fig:fig3}c, a BP sphere is placed on a semi-infinite glass substrate ($n_\textrm{sub}=1.45$) and excited by a plane wave incident from the substrate side at an oblique angle of $\theta_\textrm{inc}=42^{\circ}$. We use the f-sum corrected refractive index for BP shown in Fig.~\ref{fig:fig2}. We then perform a near-to-far field transformation to extract the scattered far-field\cite{Yang2016}. The Poynting flux of the scattered far-field in the air region is integrated over a solid angle spanning an azimuthal angle of $2\pi$ and a maximum polar angle of $\theta_\textrm{col}=\arcsin(\textrm{NA})=53^{\circ}$ to retrieve the total scattered power collected by the collection objective ($\textrm{NA}=0.8$). The total scattered power is normalized to the incident power and the geometrical cross sectional area of the particle to determine the scattering efficiency. To account for the unpolarized incident light in the experiments, we perform this simulation procedure for both transverse-electric and transverse-magnetic polarization of the incident wave. Finally, the scattering efficiency from both polarization states is averaged. The scattering efficiency simulation in Fig.~\ref{fig:fig5}c follows the same steps with the only change being that the plane wave is incident from the air side at an oblique angle of $\theta_\textrm{inc}=75^{\circ}$.

For the EELS calculations, we simulate the electron beam as an edge current with an amplitude of 1~\textmu A. The induced electromagnetic field is obtained by calculating the fields with and without the BP nanodisk in the simulation domain, and subsequently subtracting them. The energy loss is then calculated as the work rate done on the electron beam by the induced electromagnetic field~\cite{Assadillayev2021}.

\subsection*{Dark-field scattering measurements}
A custom-built inverted optical microscope was used for dark-field scattering spectroscopy of single nanoparticles (Fig.~\ref{fig:fig3}a). For the measurements presented in Fig.~\ref{fig:fig3}, the sample is illuminated from the top by a halogen lamp through a dark field condenser and the scattered light was collected by an objective (50×, $\textrm{NA} = 0.8$). For the spectrum in Fig.~\ref{fig:fig5}c, the sample was illuminated from the bottom through a dark-field objective (50×, $\textrm{NA} = 0.8$) and the scattered light was collected by the same objective. To measure the spectra, scattered light was transferred to the entrance slit of a monochromator (SpectraPro-300i, Princeton Instruments) and detected by a liquid-N$_2$ cooled CCD (Princeton Instruments). For Raman scattering measurements, the nanoparticles were excited  by a 488~nm laser (Coherent Sapphire 488–50).

\subsection*{EELS measurements and analysis}
The EELS measurements are performed in a monochromated and aberration-corrected FEI Titan operated in STEM mode at an acceleration voltage of 300~kV, providing a probe size of ${\sim}0.5$~nm and an energy resolution of 0.08~eV (as measured by the full-width-at-half-maximum of the zero-loss peak). We perform Richardson--Lucy deconvolution to remove the zero-loss peak. An EELS spectrum recorded in vacuum is used as an input for the point-spread function. Due to a small asymmetry in the zero-loss peak, the deconvolution algorithm produced an artificial EELS peak in the energy range below 0.5~eV. However, the artificial peak did not overlap with any of the observed resonances and could be safely removed using a first-order logarithmic polynomial.

The depicted EELS spectra are obtained by integrating the deconvoluted EELS data around the experimental impact parameter $b_\textrm{exp}$. This parameter is directly related to the effective radius of the particle $R$ which is extracted as a radius of the circle with the effective area of the particle found from the STEM image. The effective area of the particle is the area which is enclosed by the boundaries obtained by Otsu's thresholding of the STEM image. The integration parameter itself is calculated as a radius of the circle with the adjusted effective area of expanded/reduced initial boundaries. For the integration region centered at the nanoparticle, the experimental impact parameter changes from 0 to 0.3$R$ and denotes the nanoparticle center. For the annulus-shaped regions, the experimental impact parameter denotes the mean of the inner and outer radii with a typical radius spread of 0.3 (for example, the green region in Figure \ref{fig:fig4}(a) encloses the regions from 0.35 to 0.65$R$). The depicted EELS spectra are smoothed with a Gaussian function ($\sigma = 0.03$~eV).

The EELS maps are obtained by summing the deconvoluted EELS data in a spectral window of $0.02$~eV width centered at the resonance energies. The signal-to-noise ratio is improved by spatially binning the map, reducing the total number of pixels by a factor of 4. A Gaussian filter with $\sigma = 0.8$ pixels is applied to smooth the image.

\subsection*{Laser reshaping}
Unprocessed BP nanoparticles in acetone are irradiated with the third harmonic of a Nd:YAG laser (355~nm wavelength, 5~nm pulse width, 20~Hz repetition rate) with a fluence of 50~mJ/cm$^2$ per pulse for 10~min. 
\begin{acknowledgement}
S.~R. and A.~A. acknowledge support from the Independent Research Funding Denmark (7026-00117B). K.~S.~T. acknowledges support from the Center for Nanostructured Graphene (CNG) under the Danish National Research Foundation (project DNRF103) and from the European Research Council (ERC) under the European Union’s Horizon 2020 research and innovation program (Grant No. 773122, LIMA). K.~S.~T. is a Villum Investigator supported by VILLUM FONDEN (grant no. 37789).  H.~S. acknowledges support by JSPS KAKENHI Grant Numbers 21K14496. 

\textbf{Author contributions} M.~K.~S. constructed the computational workflow, performed the screening, developed the f-sum approach, and prepared the figures. H.~S. performed the structural and optical characterizations of nanoparticles. A.~A. performed the EELS measurements, EELS analysis, EELS simulations, and prepared a figure. D.~S. fabricated the nanoparticles and conducted laser shaping processes. M.~F. contributed to analyses and interpretation of the data and supervised the fabrication and characterizations of nanoparticles. S.~R. and K.~S.~T. conceived the idea and supervised the work. S.~R. performed the optical simulations. All authors discussed the results and contributed to the preparation of the manuscript. 

\textbf{Conflict of interests} The authors declare no competing financial interests.

\textbf{Data and materials availability} All data needed to evaluate the conclusions in the paper are present in the paper and/or the Supplementary Information. Additional data related to this paper may be requested from the corresponding author.

\end{acknowledgement}

%\begin{suppinfo}
%Supplementary Information for this paper is available online.
%\end{suppinfo}

\bibliography{achemso-demo}

\end{document}